\documentclass[a4paper]{article}  
\usepackage{graphicx}

\begin{document}
	
\begin{center}
    {\Large Geant4 FTF model description of the latest data by the NA61/SHINE collaboration
            on ${\rm ^{40}Ar+{}^{45}Sc}$ interactions }
\end{center}

\begin{center}
A. Galoyan\footnote{Veksler and Baldin Laboratory of High Energy Physics, Joint Institute for Nuclear Research,
Dubna, Moscow region, 141980 Russia},
A. Ribon\footnote{Conseil Européen pour la Recherche Nucléaire, 1211 Genèva, Switzerland},
V. Uzhinsky\footnote{Laboratory of Information Technologies, Joint Institute for Nuclear Research,    
Dubna, Moscow region, 141980 Russia}\\
\vspace{0.5cm}
on behalf of the Geant4 hadronic physics working group
\end{center}


\begin{center}
    \begin{minipage}{14cm}
        \centerline{Abstract}
        It is shown that the Geant4 FTF model, which does not include the simulation of the hard 
        parton-parton scattering and the formation of the quark-gluon plasma (QGP), describes 
        well the NA61/SHINE data on $\pi^-$ meson distributions for the interactions at 
        $\sqrt{s_{NN}}=$ 5.2, 6.1, 7.6 and 8.8 GeV. At higher energies, $\sqrt{s_{NN}}=$ 11.9 
        and 16.8 GeV, the model underestimates the data. This is considered as an indication of
        the formation of QGP at higher energies in central collisions of light and intermediate 
        nuclei than in collisions of heavy nuclei ($\sqrt{s_{NN}}\sim 6$ GeV).
    \end{minipage}
\end{center}

Recently, the NA61/SHINE collaboration has published \cite{NA61_ArSc} experimental data on $\pi^-$ meson
production in central ${\rm ^{40}Ar+{}^{45}Sc}$ interactions at beam momenta from 13A to 150A GeV/c.
The collaboration has compared their results with EPOS \cite{Epos1.99}, UrQMD \cite{UrQMD1,UrQMD2} and Hijing
\cite{Hijing1,Hijing2} model calculations. It was shown that EPOS and UrQMD models give a satisfactory
description of the data only at the momentum 150A GeV/c (see Fig.~1). At lower energies, EPOS model 
overestimates $dn/dy$ for $\pi^-$ mesons at $y\sim 0$, and UrQMD model underestimates the data. 
As it will be shown below, the Geant4 FTF model -- using the version 10.7 with minimal changes that
will be included in the next release -- describes well the data except at momenta 75A and 
150A GeV/c.

The Geant4 FTF model \cite{Geant4_3} is based on the well-known Fritiof model \cite{Fritiof1,Fritiof2}
which was widely applied for simulations of hadron-nucleon, hadron-nucleus and nucleus-nucleus interactions. 
It is assumed in the model that all nucleon-nucleon interactions are binary reactions with creation
of resonances and excited nucleons in final states. The spectrum of squared masses of the excited nucleons is 
a key ingredient of the model. We chose to parameterize the spectrum for non-diffractive interactions as 
$a/M_x^2+b$. Excited nucleons are considered as quark-gluon strings whose fragmentation produces hadrons. 
The fragmentation of strings is simulated using a modified version of the LUND algorithm \cite{LUND_PhysRep,Jetset6.2}
implemented in the Geant4 toolkit.

The Geant4 FTF model considers diffraction dissociation into high mass states, quark-exchange
processes and non-diffractive interactions. Phenomenological parametrizations of the corresponding 
cross sections are implemented in Geant4 (see details in Geant4 Physics Reference Manual 
\cite{Geant4_PRM}). For the simulation of nucleus-nucleus interactions we apply a simplified Glauber
approximation. Fermi motion of nuclear nucleons is accounted for according to the method of
Refs. \cite{Uzhi_EMU01,Khaled_Fermi}. A fine tuning of the FTF model parameters has been done using 
the NA61/SHINE data on $pp$ interactions at various energies \cite{NA61_pp}.

FTF model results for rapidity distributions of $\pi^-$ mesons in ${\rm ^{40}Ar+{}^{45}Sc}$ 
interactions are presented in Fig.~1 together with the experimental data for 0--5 \%
centralities. Having no geometrical description of the NA61/SHINE Projectile Spectator Detector 
(PSD) used for selection of central interactions, we have chosen the impact parameter range (0 -- 3.1 fm) 
to reproduce the height of the rapidity distribution at 19A GeV/c. As seen in Fig.~1, we describe quite 
well the rapidity distributions of $\pi^-$ mesons at beam momenta 13A, 30A and 40A GeV/c. At higher 
momenta we underestimate the data.

As seen also in Fig.~1, the Hijing model overestimates the data at all energies. However, a good 
description can be reached by scaling down the Hijing results by 10 -- 15 \%, though the widths of
the distributions at 13A, 19A and 40A GeV/c are overestimated. A simplest way to decrease the Hijing
results is by increasing of the probability of the one-vertex diffraction dissociation in nucleon-nucleon
interactions, as it was suggested in Ref.~\cite{Uzhi_IzvRAN}. With this, the NA61/SHINE data on
$pp$ interactions \cite{NA61_pp} are described \cite{Uzhi_IzvRAN}. There are also additional
possibilities to improve the Hijing results \cite{Khaled_SPS}.   

We did not find a simple recipe to improve the EPOS model results at low energies.

The FTF model does not consider hard interactions and quark-gluon plasma (QGP) formation. This could
explain the deviation of the FTF model results from the data at 75A and 150A GeV/c. In heavy ion collisions
the formation probably takes place at $\sqrt{s_{NN}} \simeq$ 6 -- 7 GeV (see Refs.~\cite{NA49PbPb1,NA49PbPb2}). 
Thus, we expected that the deviation of the FTF results from the ${\rm ^{40}Ar+{}^{45}Sc}$ interaction 
data would start also at $\sqrt{s_{NN}} \sim$ 6 -- 7 GeV. This is not the case. Therefore, we interpret 
our results as an indication of the QGP formation at higher energies in central collisions of light and 
intermediate nuclei with respect the collisions of heavy nuclei. It would be interesting to check this 
hypothesis by using NA61/SHINE data on ${\rm ^7Be + {}^9Be}$ interactions.
\begin{figure}[hbt]
    \centering 
    \resizebox{6in}{5.5in}{\includegraphics{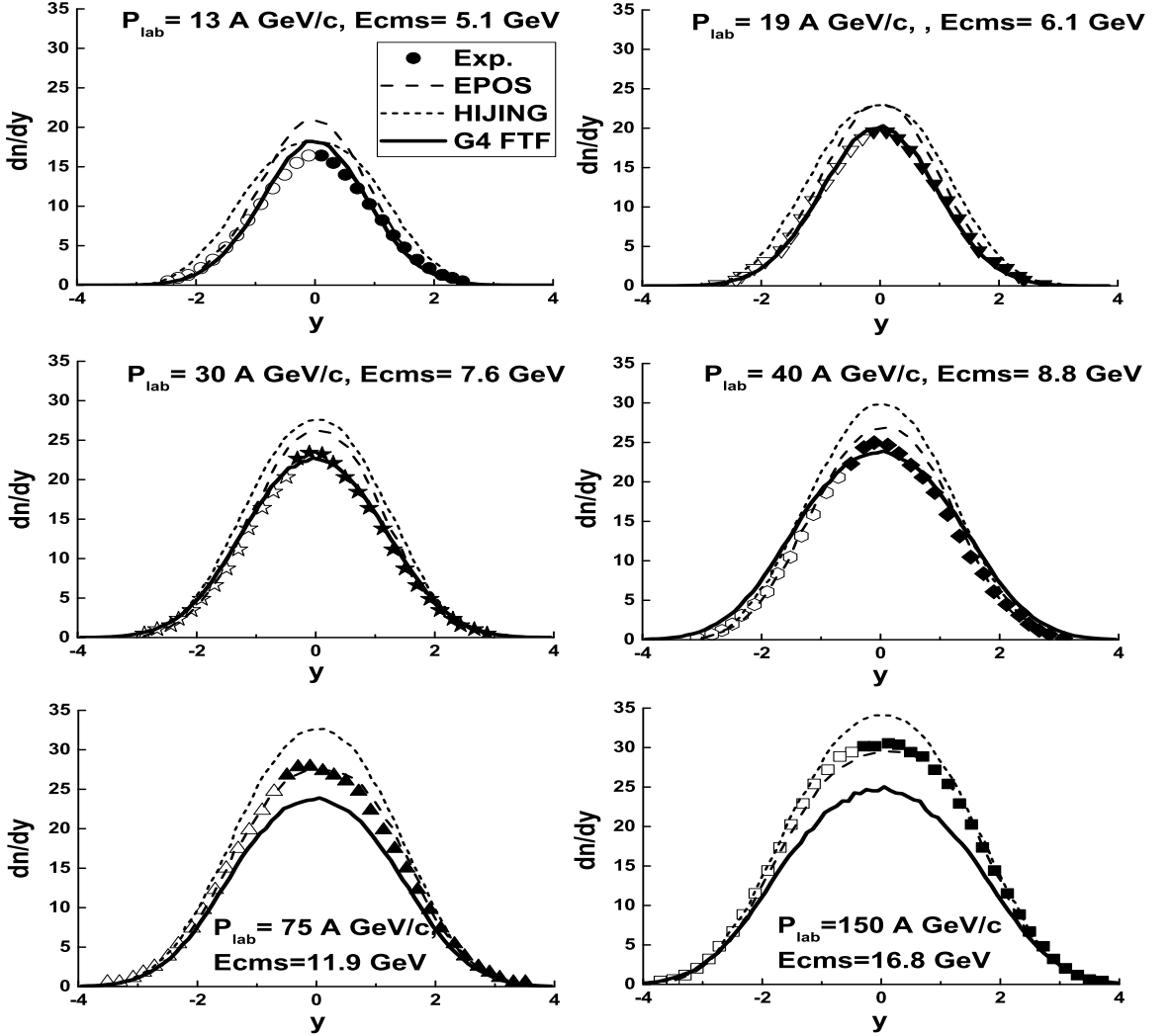}}
    \caption{Rapidity distributions $dn/dy$  of $\pi^-$ mesons for all six beam momenta. Black points 
    	are data measured by the NA61/SHINE for 0--5 \% centralities. Open points are values extrapolated
    	by the collaboration into unmeasured region. All experimental uncertainties are smaller than 
    	the symbol size. Predictions of EPOS and Hijing models obtained by the collaboration 
    	\protect{\cite{NA61_ArSc}} are shown by long and short dashed curves, respectively. The FTF model 
    	results are shown by solid lines. 
    	}
    \label{Fig1}
\end{figure}

In the following, we analyze the $p_T$ distributions of the $\pi^-$ mesons produced in the interactions.
In Fig.~2 we show the NA61/SHINE experimental data compared with model predictions.
One can see that the FTF model describes well the distributions at momenta 13A, 19A, 30A and 40A GeV/c.
The model underestimates the tails of the distributions at 75A and 150A GeV/c, and underestimates
the data at $p_T <$ 200 MeV/c. Again, this discrepancy could be caused by the absence of hard processes
in the FTF model.

The Hijing model considers hard and semi-hard interactions. But it appears that these processes are switched
on too early in the model starting from 19A GeV/c. Only at 150A GeV/c the model predictions are
in agreeement with the experimental data. Thus, it would be good to improve the energy dependence of
these processes in the model. In doing this, also the total multiplicity of produced mesons can be decreased.

For the EPOS model the situation is more complicated. The model describes well the data at 75A and 150A
GeV/c. At lower momenta, the model describes the data only at $p_T \geq$ 800 MeV/c. Thus, we believe that the
energy dependence of the hard and semi-hard processes is correctly reproduced in the model. Soft interactions
play a role at small $p_T$ ($p_T <$ 800 MeV/c) and lower projectile momenta. It is more complicated to 
improve the soft string fragmentation, as we have done in the Geant4 FTF model.
\begin{figure}[hbt]
	\centering 
	\resizebox{6in}{5.5in}{\includegraphics{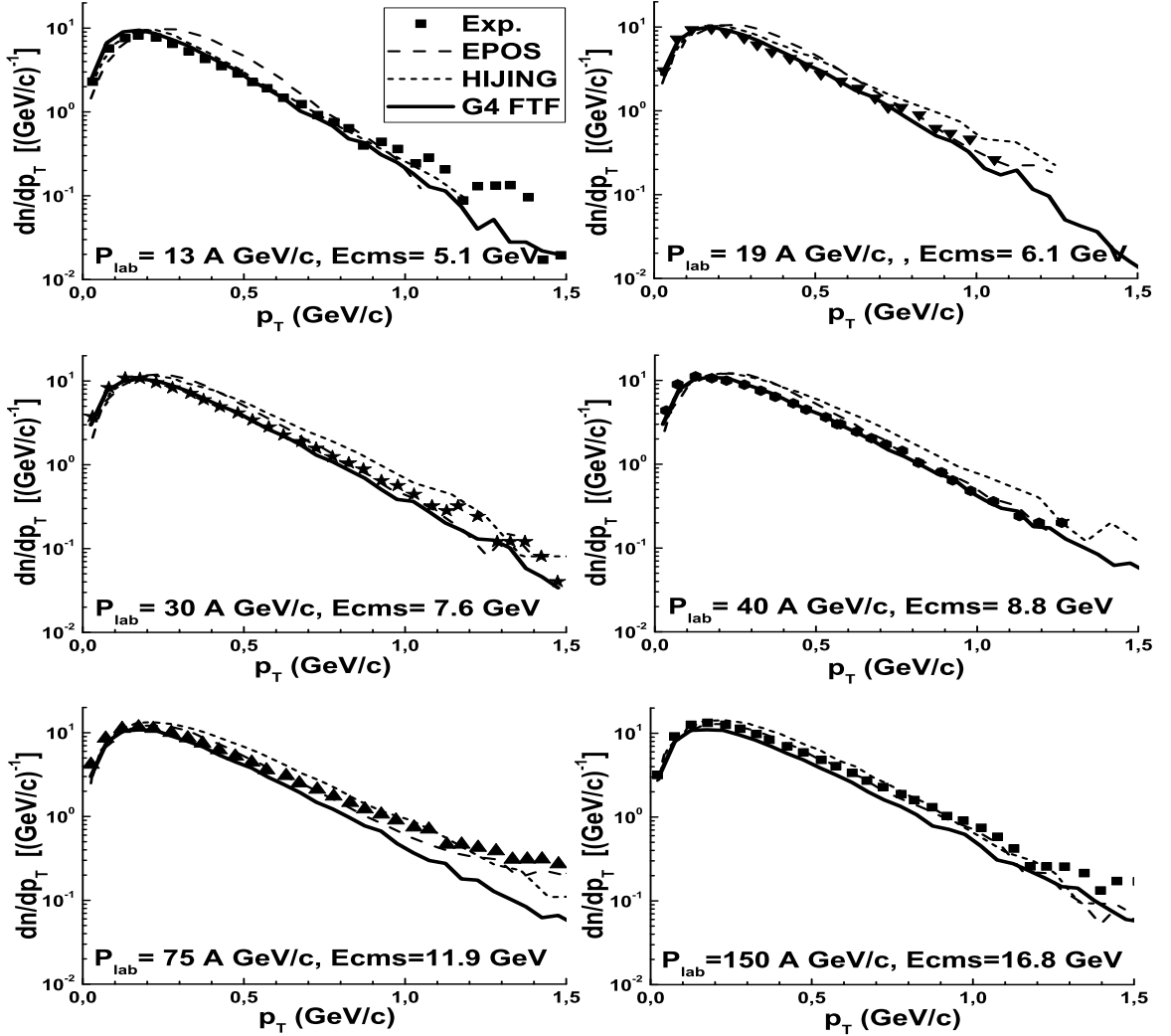}}
	\caption{Transverse momentum distributions $dn/dp_T$ at mid-rapidity ($0< y <0.1$) for all six beam momenta. 
		Points are the NA61/SHINE experimental data for 0--5 \% centralities. All experimental uncertainties 
		are smaller than the symbol size. Predictions of EPOS and Hijing models taken from
                Ref. \protect{\cite{NA61_ArSc}} are shown by long and short dashed curves, respectively.
                The FTF model results are shown by solid lines. 
	}
	\label{Fig2}
\end{figure}

\section*{Conclusion}
\begin{enumerate}
\item The Geant4 FTF model calculations are in good agreement with experimental data
of the NA61/SHINE collaboration on transverse momentum and rapidity distributions of $\pi^-$ 
mesons produced in ${\rm ^{40}Ar+{}^{45}Sc}$ interactions at $\sqrt{s_{NN}}<$ 10 GeV.
 
\item At higher energies the FTF model underestimates $dn/dy$ and the tails of $dn/dp_T$ distributions.
These discrepancies could be compensated by including hard processes and creation of QGP.

\item It would be good to reproduce the energy distributions in the Projectile Spectator Detector
of the NA61/SHINE collaboration used for centrality selection. These are presented in Ref.~\cite{NA61_ArSc}.
Their reproduction would allow to check the nuclear residual fragmentation model in FTF, and allow to use
the same criteria for centrality determination as in the experiment. It is not possible to do this at the
moment without the exact geometrical description of the detector and its response.

\end{enumerate}

A. Galoyan and V. Uzhinsky are thankful to heterogeneous computing team of LIT JINR (HybriLIT) for
support of the presented calculations. 


\end{document}